\documentstyle[12pt]{article}
\def\picture #1 by #2 (#3){
  \vbox to #2{
  \hrule width #1 height 0pt depth 0pt
  \vfill
  \special{picture #3}}}
\def\scaledpicture #1 by #2 (#3 scaled #4){{
\dimen0=#1 \dimen1=#2
\divide\dimen0 by 1000 \multiply\dimen0 by #4
\divide\dimen1 by 1000 \multiply\dimen1 by #4
\picture \dimen0 by \dimen1 (#3 scaled #4)}}
\author{Serge Galam and Alain Mauger\\
Laboratoire des Milieux D\'{e}sordonn\'{e}s et H\'{e}t\'{e}rog\`{e}nes
\footnotemark[1]\\
Tour 13 - Case 86, 4 place Jussieu, \\ 75252 Paris Cedex 05, France\\[1ex].}
\title{\hspace*{-0.7cm}Topology invariance in Percolation Thresholds}
\addtocounter{footnote}{0}
\footnotetext{Laboratoire de l'Universit\'{e}
P. et M. Curie - Paris 6, associ\'{e} au CNRS (URA n$^{\circ}$ 800)}
\addtocounter{footnote}{1}
\date{(june 3, 97)}

\begin{document}
\maketitle
\baselineskip 3.3ex
\footskip 5ex
\parindent 2.5em
\abovedisplayskip 5ex
\belowdisplayskip 5ex
\abovedisplayshortskip 3ex
\belowdisplayshortskip 5ex
\textfloatsep 7ex
\intextsep 7ex
\begin{center}
{\em PA Classification Numbers:\/} 64.60 A, 64.60 C, 64.70 P\\
\end{center}

\begin{abstract}

An universal invariant for
site and bond percolation thresholds ( $p_{cs}$ and $p_{cb}$ respectively)
is proposed. The invariant writes
$\left\{{p_{cs}}\right\}^{\frac{1}{a_s}}\left\{{p_{cb}}\right\}^{-\frac{1}{a_b}
}
 = \delta /d$ where $a_s, \  a_b$ and $\delta$ are positive constants,
and $d$ the space dimension.
It is independent of the coordination number, thus exhibiting a topology
invariance at any $d$.
The formula is checked against
a large class of percolation problems, including
percolation
in non-Bravais lattices and in aperiodic lattices as well as rigid percolation.
The invariant is satisfied
within a relative error of $\pm 5\%$ for all the twenty lattices of our sample
at $d=2$, $d=3$, plus all hypercubes up to $d=6$.

\end{abstract}
\newpage

\section{Introduction}

Percolation phenomena are active in a rather broad spectrum of physical
and non-physical problems [1]. It is now a full part
of Statistical Physics. However, most studies have been devoted
to percolation on regular
lattices, merely for convenience [2]. But even for these
lattices,
exact results are scarce. In particular site and bond percolation
thresholds are known exactly only at $d=2$ and for a few cases. Otherwise all
available thresholds are given by numerical estimates [1].

In the past several attempts were made to unify percolation thresholds.
None was really satisfactory. we have presented very recently an universal
power
law to yield both site and
bond percolation
thresholds, $p_{cs}$ and $p_{cb}$ respectively, within an excellent accuracy
[3],
\begin{equation}
p_{cs} =  p_{0s}\left\{(d-1)(q-1)\right\}^{-a_s} \ ,
\end{equation}
for site and
\begin{equation}
p_{cb}  = p_{0b} \left\{ \frac{(d-1)(q-1)}{d}\right\}^{-a_b}\ ,
\end{equation}
for bond with $d$ the space dimension and $q$ the coordination number.
The formula yields thresholds for any Bravais lattice, at
any dimension with an impressive accuracy [3].

For $d \leq 6$, two different classes were found, and identified by distinct
parameter sets
$\{p_{0i}; \ a_i\}$ where $i=s, \ b$. Both Eqs (1) and (2)
are satisfied within few per cent for all the lattices inside a given class.
The first class includes two-dimensional triangle, square
and honeycomb lattices. It is characterized by
$\{p_{0s}=0.8889; \ a_s=0.3601\}$ for
site dilution and by
$\{p_{0b}=0.6558; \ a_b=0.6897\}$ for bond dilution.
Two-dimensional Kagom\'{e}
and all
 other lattices of cubic symetry (for
$3\leq d\leq 6$) constitute the second class which is
 characterized by
$\{p_{0s}=1.2868; \ a_s=0.6160\}$ and
$\{p_{0b}=0.7541; \ a_b=0.9346\}$ for sites and bonds respectively.

Above results were obtained using a sample which includes
most common lattices mentioned above [3]. This is, however, quite
restrictive, because of the new trends in modern percolation problems
related in particular to directed percolation, percolation in aperiodic
lattices, and
rigidity percolation.

In the present work, we use a
much broader lattice sample including percolation
in non-Bravais lattices and in aperiodic lattices, and rigidity percolation as
well.
However, extension of Eqs. (1, 2) to non regular
lattices is questionable, in particular with respect to the use
of the coordination number $q$ as a relevant parameter in our universal
formula.

Indeed some lattices with equal $d$ and $q$ have different $p_c$, like,
for instance, the stacked triangle lattice and $bcc$
lattices at $d=3$ [4, 5]. In this paper we name stacked triangle the 3d-lattice
also called hexagonal lattice in the litterature, to avoid any confusion with
hexagonal compact lattice, and the 2d-triangle lattice also called hexagonal
by some authors. Moreover Eqs. (4, 5) are found not to hold
for many non-regular lattices, which is indeed not surprising.
Percolation should depend
on the degree
of lattice anisotropy, which is not included in Eqs. (1, 2) where only
the bare coordination number $q$ appears.

Therefore, at this stage, with respect to our power law,
there is no proof that, besides the dimension $d$,
there exists one
single parameter which contains all the relevant information on the
lattice topology. Yet, this parameter entering Eqs (1, 2), if it exists,
clearly
cannot be reduced to the coordination number $q$.

Nevertheless, it is worth noting that for each lattice, there exists one
single parameter $q_{eff}$ which reproduces, within an excellent accuracy,
both percolation thesholds from Eqs. (4, 5) [6]. However, at present,
we have no scheme to calculate $q_{eff}$, which then, has to be determined
from the values of
site or bond percolation thresholds.

To bypass this difficulty, we can eliminate $q$ from Eqs. (1, 2) by
combining them to obtain, for the first time in percolation theory, a
universal invariant
which combines $p_{cs}$, $p_{cb}$ and $d$.
The formula exhibits a topology invariance
and holds for
all percolation problems, including
percolation
in non-Bravais lattices and in aperiodic lattices as well as rigidity
percolation.
The invariant is satisfied, at worst
within $\mp 0.05$ for all the twenty lattices of our sample
at $d=2$, $d=3$, plus the hypercubes up to $d=6$ with the same relative
accuracy.

\section{The Invariant}

{\bf A. Non-regular lattices}\\

Let us first review the large spectum of non-regular lattices we include in our
sample
with their main characteristics.
The stacked triangle lattice is anisotropic with 6 equivalent nearest
neighbors
$(nn)$ in the $a,b$ plane (bonding angle is 60$^{\circ}$), and two non
equivalent neighbors along the $c$ axis
perpendicular to it (bonding angle is 90$^{\circ}$.
$p_{cs},\ p_{cb}$ for this lattice have been
determined very recently [4].
On another hand, the hexagonal close packed (hcp) lattice is a non-Bravais
lattice, with two
lattice sites per unit cell. The percolation thresholds of the hcp lattice
have been determined a long time ago [7].

 Aperiodic lattices are represented by
quasicrystals. Besides their own interest, these aperiodic materials with
long range order can serve as models for alloy materials, hence a growing
interest in their percolation thresholds. Our sample includes Penrose tiling
[8-10], octagonal and dodecagonal tilings [11] with chemical links
(connection via the tile edges), and ferromagnetic links (connection through
the diagonal of the tiles, which are shorter than the tile edges) [12].

The sample is also enriched by the dual of the lattices. Percolation
thresholds for the dual of quasicrystals have also been estimated in
Refs. (8-11). Those of the dual of kagom\'e, named the dice lattice are
reported
in Ref. (8). The duals of periodic 3-dimensional $fcc$, $hcp$ and
diamond lattices
have been estimated very recently [5].

We also include in the sample the case of rigidity percolation [13].
The bond percolation threshold for the existence of
stress carrying paths have been recently determined [14]
in a lattice generated by
randomly displacing the sites of a triangular lattice in dimension $d=2$. \\

{\bf B. Site versus bond percolation thresholds}\\

We can easily eliminate $q$ between the expressions of
$p_{cs}$ and $p_{cb}$ using Eqs. (1) and (2). We actually  get
the following invariant which combines
both percolation thresholds with the dimension,
\begin{equation}
\left\{{p_{cs}}\right\}^{\frac{1}{a_s}}\left\{{p_{cb}}\right\}^{-\frac{1}{a_b}}
= \frac{\delta}{d}
\end{equation}
where
$\delta \equiv
\left\{{p_{os}}\right\}^{\frac{1}{a_s}}\left\{{p_{ob}}\right\}^{-\frac{1}{a_b}}$
,

Our above formula shows for the first time a topology invariance with
respect to percolation
thresholds. To check its validity  against our sample of lattices,
it is more convenient to rewrite it as
\begin{equation}
p_{cs} = \delta^{a_s}
\left\{ d^{-a_b} p_{cb}\right\}^{\frac{a_s}{a_b}}\ ,
\end{equation}
to have a better graphic representation.

We have plotted in Fig. (1)
$\log (p_{cs})$ vs $\log(d^{-a_b} p_{cb})$.
After Eq. (3), the universal curve in Fig. (1) reduces to a straigt line.
The agreement of Eq. (3) with the data is
impressive for all the
lattices, with the exception of the dual of diamond, which
is far from the straight line in Fig. (1).

Percolation thresholds decrease with increasing space dimensionality.
Therefore, the best test for the law when $d$ varies from 2 to 6 is provided
by the relative deviation $\Delta p_c/p_c$, since the absolute deviation
$\Delta p_c$ is necessarily small at large $d$. That is why we have plotted in
Fig. (1) the logarithm of the quantities of interest, in linear scale,
instead of
the quantities in a logathmic scale. The deviation from the straight line
in Fig. (1) is thus a measure of the relative deviation from the law. Also,
the accuracy
of Eq. (2) is evident from the correlation coefficient,
\begin{equation}
r = \frac{n \sum xy - \sum x \sum y}{\sqrt{[n \sum x^2 - (\sum x)^2][(n
\sum y^2 -
(\sum y)^2]}}
\end{equation}
on the data $(x,y)$ on the $n=9$ lattices of the first class and the $n=13$
 lattices
of the second class (once the dual of diamond has been excluded) in a linear
regression analysis. The result is $r=0.997$ for the first class, and
$r=0.9994$
for the second class. Yet the smaller value of $r$ in the first class may be
due to the lack of accuracy in the determination of the percolation thresholds
for the octa- and dodecagonal quasicrystals and their duals. The associated
numerical estimate is reported with only two decimals for bonds,
and three decimals for sites [10], against four decimals in most
other cases.

The parameters $p_{0s}, p_{0b}$, $a_s$ and $a_b$ have been
determined from the fit of Eqs (1) and (2) [3]. However, these equations are
only approximate, so that the values of $a_s$ and $\delta$ deduced from this
work is not the best choice for these
parameters entering Eq. (3) or Eq. (4). Instead, we have determined
$a_s$ and $\delta$ independently, from the least square fit of the data in
Fig. (1) by a straight line. The same linear regression analysis which has
provided
us with the correlation coefficient $r$ above mentioned gives
\begin{equation}
\left\{\begin{array}{ll}
a_s = 0.3670; \delta = 1.3638 & \mbox{for the first class}\\
a_s = 0.6068; \delta = 1.9340 & \mbox{for the second class}\\
\end{array}
\right.
\end{equation}
We have listed in Table (1) the quantity
\begin{equation}
C=\frac{\delta}{d (p_{cs}^{\frac{1}{a_s}} p_{cb}^{-\frac{1}{a_b}})}
\end{equation}
The relative error with respect to Eq. (3) for each lattice is then
$\mid C-1 \mid$.

Though the deviations are small, they are significant,
showing our formula is not exact. One consequence is the inconsistency in the
numerical value of $a_s$ in Eq. (6), different from the value of $a_s$
deduced from the least-square fit of Eq. (1) in ref. [3] in the case of the
second class. In particular,
we have checked that this difference is not solely related to the addition of
many lattices to the initial sample in [3] for both classes: when all the
lattices in the present work are taken into account, $a_s$
deduced from the fit of Eq. (1) is 0.618 for
the second class, close to 0.616 reported when the samples are reduced to
the initial set of basic lattices [3]. The difference with 0.6068 in Eq. (6)
is then due to the fact that none of the Eqs (1,2), and (4) are exact.

Actually, from Eqs. (7), an upper limit of the
relative error is given by
\begin{equation}
\mid C-1 \mid \leq \frac{1}{a_s} \frac{\Delta p_{cs}}{p_{cs}} +
\frac{1}{a_b} \frac{\Delta p_{cb}}{p_{cb}}  .
\end{equation}
It is reached when
deviations of $p_{cs}$ and $p_{cb}$ from Eqs (1,2) are not correlated. For the
second class, as an example, the relative accuracy of Eq.(1) according to
reference [3] is $\Delta p_{cs}/p_{cs}=3\%$, while that of Eq. (2) is
$\Delta p_{cb}/p_{cb}=2\%$, hence $\mid C-1 \mid \leq 7\%$.
According to Table 1, the deviation of $C$ from unity is better than expected,
namely within 5 per cent. This accuracy for the universal law
in Eq. (3) means a correlation between deviations of Eqs (1) and (2), so that
Eq. (3) can be satisfied, despite Eqs. (1) and (2) are not.
The most outstanding illustration is provided by the
rigidity percolation case. Here, Eq. (3) is satisfied, with $C$= 0.97, but the
site and the bond percolation
thresholds according to Eqs. (1) and (2) are those expected for a lattice with
$q=3$, while the actual coordination number is $q=6$.

The topology invariance law in Eq. (3) is satisfied (within 5\%) for all the
lattices
with the exception of the dual of diamond. We then conclude that the dimension
$d$ which is the only variable in this equation is a robust parameter.
On another hand, the percolation thresholds of some
lattices are significantly different from those expected from Eqs. (1) and (2).
In a prior work [15], we had already given arguments to substitute $q$ by
$(q-1)$ in formulas to determine percolation thresholds in frustrated lattices.
 The rigidity percolation with $q_{eff}=3$ against
$q=6$ is another evidence
that the coordination number is not a robust parameter.
This is a limit to the application of
Eqs (1) and (2) as long as any model to determine $q_{eff}$ [3]
from the topology
of the lattice is lacking. Eq. (3) is then a significant improvement, as it
does not involve any additional unknown parameter.
Moreover, this law is the first link between site and bond percolation
thresholds.

We have already argued [3] that Eqs. (1, 2) violate the $(q-1)^{-1}$ expansions
for the d-dimensional simple hypercubic lattice percolation thresholds, and
do not match the Bethe asymptotic limit. This is also
the case for Eq. (3). Actually, in the large $d$ limit, the leading term
for both percolation thresholds should be the Bethe term $p_c^s \sim p_c^b
\sim 1/(q-1) \sim 1/(2d)$. After Eq. (4) this limit requires:
\begin{equation}
\frac{1}{a_s}-\frac{1}{a_b} = 1; \delta = \frac{1}{2}
\end{equation}
The deviation from our results in Eq. (6) gives evidence that our formula
in Eqs. (3, 6) is not compatible with exact 1/d expansion. We then expect that
the range of validity for our farmula is the same as for Eqs (1, 2), namely
$d\leq 7$, after Ref. 3.

\section{Conclusion}

 Lattices with a higher
coordination number have lower percolation thresholds [5]. Yet there are
exceptions like
the Kagom\'e lattice at $d=2$ and the dual of diamond at $d=3$.

At $d=2$ there are theoretical arguments according to which
the bond percolation threshold of a lattice and its dual should add to
one. However, no regularity has been found in three dimensions [5].

The universal law provided by Eq. (3) is then the first relation which
links site and bond percolation thresholds.
It applies in any dimension
up to $d=6$, and extends to any kind of lattice.
Departure from this universal law for all the
lattices in any dimension is within few per cent.

In addition, the robustness of our universal law supports the extension
to more complex problems such as
rigidity percolation. We then expect it to be satisfied for any
percolation problem, with very few out-liers.
The only one we have found so far
is the dual of diamond for which indeed an anomalous site percolation threshold
has been reported.


\subsection*{Acknowledgments.}
We would like to thank Monsieur Dietrich Stauffer
for stimulating vibrations.

\newpage
\vspace{3.0cm}
{\LARGE References}\\ \\
1. {\sf M. Sahimi}, {\em Application of Percolation Theory} (Taylor
and Francis, London, 1994). \\
2. {\sf D. Stauffer and A. Aharony}, {\em Introduction to Percolation
Theory}, 2nd ed. (Taylor and Francis, London, 1994). \\
3. {\sf S. Galam and A. Mauger}, Phys. Rev. E \underline {53}, 2177 (1996). \\
4. {\sf Van der Marck}, Phys. Rev. E \underline {55}, 1228 (1997);
{\sf S. Galam and A. Mauger}, Phys. Rev. E \underline {55}, 1230 (1997). \\
5. {\sf Van der Marck}, Phys. Rev. E {\em in press}. \\
6. {\sf S. Galam and A. Mauger}, Phys. Rev. E, in press. \\
7. {\sf V.K.S. Shante and S. Kirpatrick}, Adv. Phys. \underline {20}, 326
(1971). \\
8. {\sf J. P. Lu and J. L. Birman}, J. Stat. Phys. \underline{40}, 1057
(1987).\\
9. {\sf F. Babalievski and O. Pesshev}, C. R. Acad. Sci. Bulg. \underline{41},
85 (1988).\\
10. {\sf F. Yonesava, S. Sakamoto, K. Aoki, S. Nose and M. Hori}, J.
Non-Cryst. Solids
\underline{123}, 73 (1988).\\
11. {\sf F. Babalievski}, Physica \underline{A220}, 245 (1995).\\
12. {\sf D. Leduie and J. Teillet}, J. Non-Cryst. Solids \underline{191},
216 (1995).\\
13. {\sf C. Moukarzel and P.M. Duxbury}, Phys. Rev. Lett., in press. \\
14. {\sf D. Jacobs and M. F. Thorpe}, Phys. Rev. Lett., in press. \\
15. {\sf S. Galam and A. Mauger}, Physica \underline{A205}, 502 (1994).

\newpage
{\LARGE Figure captions}\\ \\

Fig. 1. Decimal logarithm of site versus bond percolation thresholds.\\

\newpage

\begin{table}

\label{tbl}
\begin{center}
\hspace*{-2.5cm}
\begin{tabular}{|c|c|c|c|}
\hline
\multicolumn{4}{|c|}{first class}\\
\hline
\multicolumn{1}{|c|}{Lattice}
&\multicolumn{1}{|c|}{$p_c$ (site)}
&\multicolumn{1}{|c|}{$p_c$ (bond)}
&\multicolumn{1}{|c|}{$C$}\\
\hline
Square & 0.5928 & 0.5 & 1.05\\
Honeycomb & 0.6962 & 0.6527 & 0.99\\
Triangular & 0.5 & 0.34729 & 0.98\\
Rigid perco. & 0.6975 & 0.644 & 0.97\\
Dice & 0.5851 & 0.476 & 1.01\\
Penrose & 0.5837 & 0.477 & 1.02\\
Octa.chem.links & 0.585 & 0.48 & 1.01\\
Octa.ferro.links & 0.543 & 0.40 & 0.98\\
Dode.chem.links & 0.628 & 0.54 & 1.00\\
\hline
\hline
\multicolumn{4}{|c|}{second class}\\
\hline
\multicolumn{1}{|c|}{Lattice}
&\multicolumn{1}{|c|}{$p_c$ (site)}
&\multicolumn{1}{|c|}{$p_c$ (bond)}
&\multicolumn{1}{|c|}{$C$} \\
\hline
Kagom$\acute{e}$ & 0.6527 & 0.5244 & 0.99\\
dual of Penrose & 0.6381 & 0.5233 & 1.02\\
Dode.ferro.links & 0.617 & 0.495 & 1.02\\
hexag. compact & 0.204 & 0.124 & 0.96\\
stacked triangle & 0.2623 & 0.1859 & 0.98\\
Diamond & 0.43 & 0.1859 & 0.95\\
simple cubic & 0.3116 & 0.2488 & 1.00\\
bcc & 0.246 & 0.1803 & 1.05\\
fcc & 0.198 & 0.119 & 0.96\\
dual of fcc & 0.3341 & 0.2703 & 0.98\\
dual hexag. comp.& 0.3101 & 0.2573 & 1.05\\
dual of diamond & 0.3904 & 0.235 & 0.66\\
sc (d=4) & 0.197 & 0.1601 &1.00\\
sc (d=5) & 0.141 & 0.1182 & 1.00\\
sc (d=6) & 0.107 & 0.0942 & 1.03\\
\hline
\end{tabular}
\end{center}

\caption{\sf exact estimates of percolation thresholds $p_c$ and universal
constant $C$ defined in Eq. (6). Deviation of $C$ from unity measures
the departure from the invariant.
 All the lattices belonging to the first class are in dimension $d=2$.
Those of the second class are in $d=2$ (the three first ones), $d=3$
(next nine lattices) and $d=4, 5, 6$ for the simple hypercube (sc).}
\end{table}

\end{document}